# A biological sequence comparison algorithm using quantum computers

**Büsra Kösoglu-Kind,** Institute of IT Management and Digitization Research (IFID), FOM University of Applied Sciences in Economics and Management, 40476 Dusseldorf, Germany

**Robert Loredo,** International Business Machines Corporation (IBM), Armonk, NY, 10504, USA

**Michele Grossi,** European Organization for Nuclear Research (CERN), Geneva 1211, Switzerland

**Christian Bernecker,** International Business Machines Corporation (IBM), 80339 München, Germany

**Jody M Burks,** International Business Machines Corporation (IBM), Armonk, NY, 10504, USA

**Rüdiger Buchkremer\*,** Institute of IT Management and Digitization Research (IFID), FOM University of Applied Sciences in Economics and Management, 40476 Dusseldorf, Germany

# Abstract

**Genetic information is encoded as linear sequences of nucleotides, represented by letters ranging from thousands to billions. Differences between sequences are identified through comparative approaches like sequence analysis, where variations can occur at the individual nucleotide level or collectively due to various phenomena such as recombination or deletion. Detecting these sequence differences is vital for understanding biology and medicine, but the complexity and size of genomic data require substantial classical computing power.**

**Inspired by human visual perception and pixel representation on quantum computers, we leverage these techniques to implement pairwise sequence analysis. Our method utilizes the Flexible Representation of Quantum Images (FRQI) framework, enabling comparisons at a fine granularity to single letters or amino acids within gene sequences. This novel approach enhances accuracy and resolution, surpassing traditional methods by capturing subtle genetic variations with precision.**

**In summary, our approach offers algorithmic advantages, including reduced time complexity, improved space efficiency, and accurate sequence comparisons. The novelty lies in applying the FRQI algorithm to compare quantum images in genome sequencing, allowing for examination at the individual letter or amino acid level. This breakthrough holds promise for advancing biological data analysis and enables a more comprehensive understanding of genetic information.**

# 1   Introduction

According to Lawrence et al., there are ongoing significant international endeavors to develop comprehensive gene catalogs that can identify the genes responsible for the onset and progression of diseases[1]. Recent findings suggest that the list of cancer-associated genes deemed significant is expanding rapidly and in a manner that seems improbable. Wang et al. emphasize the urgent requirement for practical diagnostic tools to address the COVID-19 pandemic[2]. They note that the current targets, particularly the nucleocapsid (N) gene primers, and probes widely used for diagnosis, are experiencing mutations. Hasin et al. argue that the sequence of amino acids not only plays a crucial role in genomics but also other "omics" such as proteomics, transcriptomics, and metabolomics[3]. Amino acid sequences and mutations provide the underlying information flow contributing to disease development. In bioinformatics, amino acid sequence information is represented as letter sequences, which can be analyzed to identify differences and commonalities in RNA and protein gene products. This analysis helps uncover functional and structural insights. Genome analysis is also vital for optimizing industrial drug or food production processes. Quantum computers, a revolutionary technology, promise to advance sequence research in medicine and biochemistry. Limitations of the current near-term systems include but are not limited to decoherence, gate fidelity, connectivity, and lack of error correction [4].

Quantum computing has made significant contributions to various biochemical inquiries. Fox et al. investigate the use of Quantum Annealers (QAs) to predict the secondary structure of RNA[5]. They demonstrate the speed and efficacy of QA in identifying low-energy solutions, presenting a competitive alternative to classical algorithms and offering potential advancements in RNA folding predictions. Wong and Chang propose a quantum algorithm utilizing Grover's search algorithm for protein structure prediction. Their approach achieves a quadratic speedup compared to classical methods, showcasing the potential of quantum computing to enhance the efficiency and accuracy of protein structure prediction[6,7]. Successful simulations on IBM Quantum's qasm simulator support their findings, emphasizing the importance of this advancement in drug and vaccine development. Robert et al. propose a resource-efficient quantum algorithm for protein folding[8], while Chandarana et al. introduce a digitized-counter-diabatic quantum algorithm for the same purpose[9]. These algorithms demonstrate significant speed improvements over classical approaches, offering promising avenues for protein folding research. Nałęcz-Charkiewicz and Nowak propose an algorithm for DNA sequence assembly using quantum annealing, showcasing its superior speed compared to classical methods[10]. Boev et al. present a genome assembly algorithm that combines quantum and quantum-inspired annealing, achieving notable speed enhancements over classical algorithms[11]. Sarkar et al. propose a quantum algorithm for de novo DNA sequence reconstruction based on the Variational Quantum Eigensolver (VQE) algorithm, showing a significant speed advantage over classical algorithms[12]. These studies collectively indicate that quantum computing holds immense potential for revolutionizing protein folding and DNA sequence assembly.

However, no study has specifically addressed the search for single amino acid mutations at the letter level, highlighting an area that requires further exploration. Quantum computing offers potential acceleration for solving complex problems, including mutation searches and pattern recognition in gene sequences[13]. Unlike classical computers that use single bits with two possible states (0 or 1), quantum computers utilize quantum mechanical states, such as superposition, entanglement, and inference, to find solutions [14,15]. These principles have demonstrated speed-ups in various problem domains. They operate using qubits, the fundamental units similar to classical bits. However, quantum systems are still in their early stages of development and have not reached the point of quantum advantage, where they can solve classically intractable real-world problems. Limitations of near-term quantum systems include decoherence, gate fidelity, connectivity, and lack of error correction[4]. Efforts are focused on optimizing quantum circuits to minimize errors through software and hardware transpiration techniques. These techniques help mitigate decoherence and maintain the quantum state during computations[16]. Hardware measures ensure the isolation of qubits from environmental interferences, presenting a scientific challenge to minimize error sources. Quantum computers hold potential beyond life sciences, extending to finance and manufacturing. Our paper aims to identify sequence similarities using a quantum computer based on existing encoding patterns and techniques[17,18]. Image processing, a well-researched area in computer science, is chosen as an approach for mutation searches due to its similarity in identifying sequential digital information. The implementation utilizes a Flexible Representation of Quantum Images (FRQI) to compare quantum images on an IBM quantum computer[19]. This method effectively encodes differences among data using quantum states.

In genome sequencing, one of the known challenges is the search for exact or approximate matches[20]. String matching algorithms are commonly employed for the former, while the Levenshtein distance is used for the latter[21]. These algorithms fall under the category of alignment approaches, which can be further classified as global, local, or heuristic alignments. Notable alignment methods include the Needleman-Wunsch algorithm[22], Smith-Waterman algorithm, BLAST[23], and FASTA[24]. However, state-of-the-art techniques are still characterized by high costs and time requirements. The Needleman-Wunsch algorithm performs a recursive calculation of two sequences within a matrix. One sequence is represented along the x-axis, while the other is along the y-axis. The similarity between the sequences is evaluated using a scoring function that penalizes gaps and mismatches. This algorithm has a time complexity of $O(mn)$, where m and n denote the sizes of the two sequences. The memory consumption is $O(nm)$. For instance, comparing two sequences of size 100,000 using a 4-bit integer representation would result in a memory consumption of 37 gigabytes[22]. To mitigate this, the Hirschberg algorithm reduces memory consumption to linear space, specifically $O(m + n)$[25] [26]. Quantum memory provides an alternative approach, where information can be stored in the superposition of qubits. The entanglement of qubits enables exponential memory growth. In a one-dimensional sequence, only $[\log_2(n)]$ qubits are required, whereas a classical system would need N bits [9] [10] (refer to Figure 1). When comparing two sequences, the memory consumption is given by ($[\log_2(n)] + \log_2(m)$) qubits.

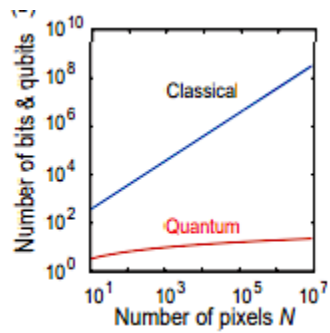

**Figure 1: Comparing the two states.** FRQI is used for representation with a complexity of $O(2^{4m})$. The amplitude estimation has a complexity of $O(\sqrt{n})$, and 1D Hadamard wavelet transform takes time of $O[poly(m)]$[27][28]. The comparison of each image can be done in $O(n^2)$.

**Understanding the abbreviated language of biological information.** Biological information, such as genetic sequences, is represented by letters denoting chemical residues in the order of their appearance in the biochemical polymer chain. In nucleic acid sequences (DNA, RNA), the individual nucleotides are indicated by phosphate groups attached to the sugar's 5' carbon (the "5' end") and unmodified hydroxyl groups on the sugar's 3' carbon (the "3' end"), following a 5' to 3' directionality. In protein sequences, the representation starts with an amino acid having a free amino group and ends with an amino acid having a free carboxyl group (N to C direction). Each type of biological sequence has its specific letter code representation, such as A, G, C, and T for DNA nucleotides (Adenine, Guanine, Cytosine, and Thymine), and amino acid codes like W for Tryptophan, G for Glycine, P for Proline, and so on. [29].

**Biochemical identity or similarity.** The most common method for comparing biological sequences involves comparative studies using sequence alignments. In alignments, sequences are arranged so that regions of importance can be compared (see Figure 2) and evolutionary, functional, structural, or other features can be identified and analyzed [30]. At least two sequences are required for the simplest of alignments (pairwise alignments comparing two sequences), and multiple sequence alignments compare three or more sequences. Common themes analyzed for various reasons (mutational analyses, phylogenetic relationships, comparative sequence analysis for structure-function relationships [31] include identity and similarity. Identity between sequences is where, in an alignment, two or more sequences contain the same exact residue (e.g., adenine and adenine) at the same relative position in that residue sequence. Sequence similarity means that the residues are from the same biochemical family. In nucleotides, purines are similar (adenine and guanine), and pyrimidines are similar (cytosine and thymine). In amino acids, similarity can be informed by the makeup of the amino acids R groups. Identity and similarity scores for a given alignment are calculated using similarity matrices such as BLOSUM or PAM [32]. Because we mention two different contexts for identity and similarity (biochemical versus frequency), sequence identity and similarity about biochemical information (Adenine, Guanine, etc.) will be referred to as "biochemical" identity or similarity.

```
         SEQUENCE.1   1    AGCTGTTACCTATGC   15
A                         ||||  ||  ||||||
         SEQUENCE.2   3    AGCTACTA-CTATGC   16

         SEQUENCE.A        AGCTGCTAGCTGTGC
         SEQUENCE.B        AGCTGCCA-CTATGC
B        SEQUENCE.C        AGC-GCTAGCTATGC
         SEQUENCE.D        AGCTGCCAGCTACGC
                                                    G G
                                                   A   C
         Pairingmask       -111----111-            U-A
         SEQUENCE.A        aGGUaggcACCu            G-C 10
C        SEQUENCE.B        aACCaggcGGUu            G-C
         SEQUENCE.C        aGUGagucCGCu           1A   U
         Support           -SSS----SSS-         "SEQUENCE A"
                                                   Helix 1
```

**Figure 2. Types of sequence alignments.** (**a**) Pairwise sequence analysis showing residues biochemically identical between alignment positions. Relative positions are compared vertically by column, with individual sequences on each row. Biochemically Identical residues are indicated with "|." Position numbers on either side of each sequence indicate position numbers in the original sequence record. This is similar to the output of the BLAST algorithm [23]. A Gap, or section of the sequence for which there is no information, is indicated by a. "-." The gap may signal a deletion mutation for homologous or evolutionarily related sequences. (**b**) A simple multiple sequence alignment from Clustal [33]. Sequence positions are compared vertically, with gaps introduced (with a penalty) to move segments of the sequence string in a manner that maximizes the alignment score calculated by a given algorithm. (**c**) A comparative sequence alignment for identifying secondary structure features in RNAs [34,35]; Sequences are arranged in phylogenetic or evolutionary order. Then a proposed location marker indicating features such as base pairing (e.g., a pairing "mask") is placed at the top of the alignment. In this example, proposed canonical Watson-Crick and G-U "wobble" pairs are indicated in upper case, and mismatches or non-pairs are indicated in lower case. The 2:1 Larsen-Zwieb rule is the benchmark for identifying compensatory base changes or changes between sequences resulting in a base pair. The base pairing supported ("S") by this analysis is reported at the bottom of the alignment for each position of the helix and shown on the right with position numbers in a secondary structure diagram.

## 2 Results

**Mapping genetic sequences onto quantum computers using Toffoli and basis gates.** In current quantum computing algorithms, information is generally mapped to a quantum state that represents the data in a way to be able to run on a quantum computer. In this case, the quantum state will be encoded via gate-based operations provided by the quantum computer. To map a genetic sequence on a quantum computer, we use a set of gates to represent two pieces of information: the position of the biological sequence and the value at that position. We use multi-control gates to entangle the information together on a quantum computer. One of the most common multi-control gates is the Toffoli gate. In Figure 3 below, a Toffoli gate [36], which in our experiment entangles three qubits together where the first two represent the position and the third is the value at the position. A Toffoli gate is a 3-qubit gate with two control qubits and one target qubit. In Figure 3, the Toffoli gate is shown with three qubits. q0 and q1 represent the control qubits and influence the target qubit q2. The control, identified by the solid sphere, triggers the action of the target qubit identified by the larger sphere with a symbol indicating the type of control; in this case, it is a NOT gate. When both q0 and q1 are enabled (set), the action at the target qubit is performed. If q0 and q1 are not enabled, then no action is performed on the target qubit, q2. Quantum gate-based systems have one similarity to classical techniques in that they use what is referred to as basis gates to construct more enormous complex gates, such as the Toffoli gate. In classical systems, these basis gates are often referred to as universal gates, such as the AND, NOT, and NAND gates. Quantum systems also have basis gates, U and CNOT gates are two examples of single-qubit and multi-qubit gates, respectively.

Therefore, to construct a Toffoli gate using the basis gates available on a quantum computer, you will need a combination of nine universal gates, also called U-gates, and six multi-qubit gates (CNOT), which when combined, as illustrated in Figure 3, a circuit depth of 11. A qubit can execute several such gates, which initially do not involve restrictions. But the more gates applied on a qubit, the deeper the quantum circuit becomes. Over time will begin to experience some of the effects of noise, such as decoherence, which could then introduce errors to the results of your experiment.

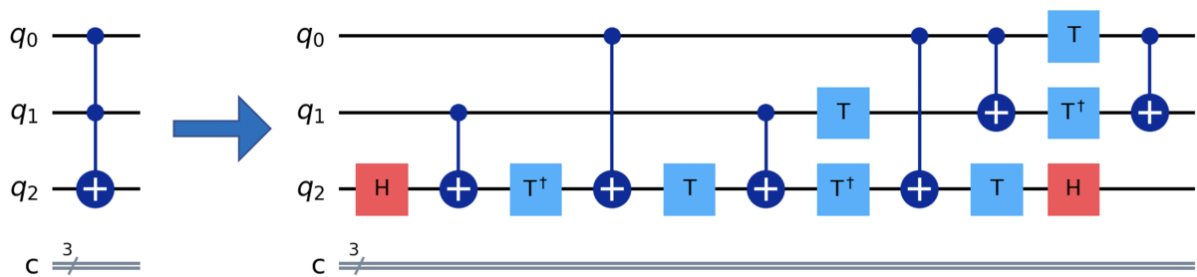

**Figure 3. A Toffoli gate or CCX** (left) and its decomposition is represented in hardware native 1-qubit, 2-qubit gates (right). Despite its simple ideal circuit representation, the Toffoli gate and its multi-dimension extension (MCX) are largely adopted in this work, resulting in long circuit decomposition.

**Representing DNA residues on a quantum circuit through parameterized rotations.** Conveying the information onto a quantum computer involves a few steps. The first step is to determine to represent the states of our sequence. The table below shows four DNA residues and, thus, a gene sequence defined on a quantum circuit where the theta function will represent the four variations. It is required to parameterize the quantum states from angles to a different value in the code because of the software convention. Rotations around the axis, i.e., π, are used as a qubit state. This allows us to represent each state by an angle, represented in degrees or radians. These angles determine the position in which the respective qubits are rotated so that they can be recognized directly based on the angle definition, whether the respective qubit is encoded as A, C, T, or G. The angle definition is used to determine the qubit's state.

| DNA Residues | Quantum state representation | Parameterized gate rotations |
|---|---|---|
| Adenine (A) | $\pi$ (180°) | $\pi/4$ |
| Cytosine (C) | $\pi/2$ (90°) | $\pi/8$ |
| Tyhmine (T) | $\pi/6$ (30°) | $\pi/24$ |
| Guanine (G) | (0°) | 0 |

**Table 1: Quantum state representations** of DNA residues and parameterized U gate rotations used in this study.

The qubit represents adenine (A) by setting the parameter to π. The Multi-control qubit gate definition of π derived from mapping the state to the basis gate on the device. Basis gates are native to the quantum hardware commonly used to create more complex gates, like universal gates in classical computing. To better illustrate this, Figure 4 represents the Adenine in a circuit. The five wires represent the qubits on which the multi-control gates and operators are applied. The first qubit, titled "strip," illustrates how we will indicate which sequences we wish to compare; this will be explained in more detail in the next section. The following figure represents the encoding of the sequence at position (1,1) with a value for 'A' (Adenine), which is represented as a quantum state of π, which is parameterized as a π/4 rotation.

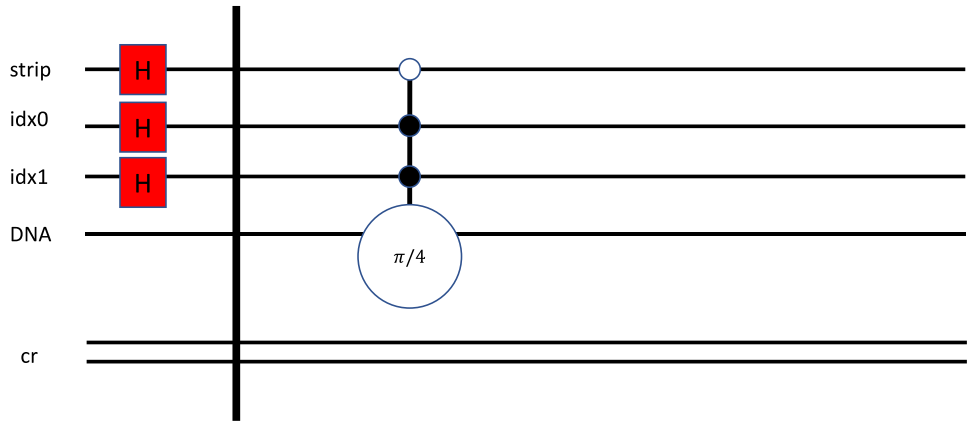

**Figure 4. According to the encoding strategy adopted, Adenine representation as a quantum circuit** represents the multi-control gate CCCNOT decomposition.

In this work, in comparison to the Adenine representation in Figure 4, Thymine is represented in Figure 5 with a rotation of π/6. It is defined on the circuit as π/24 because of the multi-qubit gate definition. After encoding all values, we add a Hadamard gate to the strip qubit, followed by a measurement operator that will measure the value of the strip qubit, which we will then use to calculate the similarity.

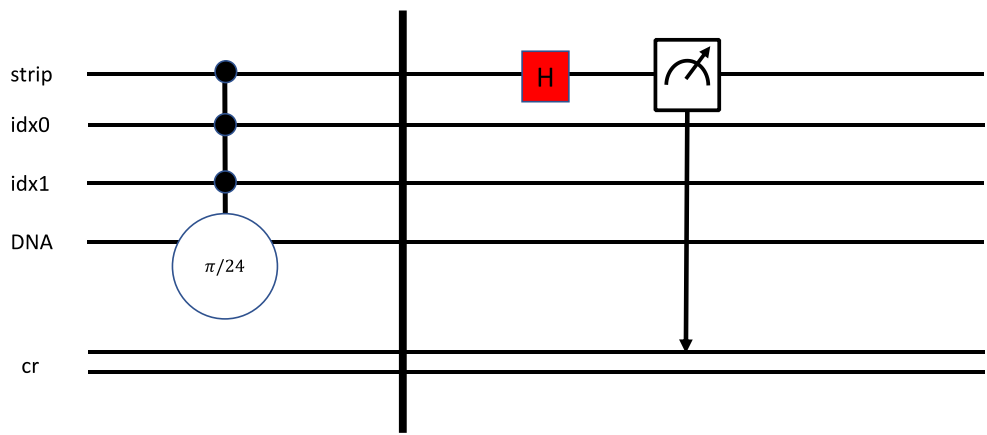

**Figure 5. Thymine representation as a quantum circuit**, representing the decomposed circuit.

**Sequence comparison with Quantum.** This section introduces and describes a quantum algorithm implementation for sequence comparison. Sequence comparison is vital for identifying functional regions, mutations such as polymorphisms, determining different forms of genes such as alleles that result in specific traits or diseases, and many other techniques [37]. In the population, different alleles exist that lead to different expressions in the individual's phenotype and ultimately result in e.g., brown or blond hair. According to the Human Genome Project [38], it is easier to identify mutations that cause a particular disease, leading to improved diagnoses, prevention, or therapies. One of the latest and most promising techniques is the CRISPR method of the two Nobel Prize winners, Jennifer Doudna and Emmanuelle Charpentier, also called gene scissors, that promises new possibilities against cancer, AIDS, and several hereditary diseases [39].

We compare two sequences in a pairwise alignment at the quantum implementation to detect patterns and mutations as in classical algorithms such as Needleman-Wunsch and Smith-Waterman; the two sequences are compared position by position [22,40].

The "similarity" approach and the technique described by Fei Yan et al. [41] compare two gene sequences, which analyzes the sequence information using the strip qubit to identify which sequence pairs to reach. The method includes an evaluation of the similarities between the encoded quantum sequence representation of the same size, here replaced by the four nucleotides. A similarity value is estimated based on the probability distribution of the readouts from quantum measurements. The proposed method provides a significant speed-up compared to traditional computers as it requires less computational power [41]. This is due to the use of various quantum gates to transform all the information-encoded sequences into the strip. This is done by first preparing the sequences into quantum states where each value contains the index (position) in the sequence and is assigned a variable that includes A, C, T, and G. In this experiment, we applied a 2x2 matrix to represent a subset of the gene sequences, which are then compared to each other using a single qubit which we will refer to as a strip qubit [41]. The reference sequence we use to compare against different sequences will be specified by the strip0 labeled qubit.

In this example, the Adenine is represented on all four entries on the circuit, as illustrated in Figure 6. A possible encoding strategy adopted here to map letters, the nucleic bases, is to represent all four possible positions 00, 01, 10, and 11 on the quantum circuit. Our system has a matrix with positions labeled 00, 01, 10, and 11. The first reference indicates the encoding for the next strip or sequence. The index for the following reference, strip 0, can be 0 or 1. If it is 0, it represents the reference image, whereas if it is 1, it represents the compared sequence image. This distinction between the two is necessary to avoid confusion. The positions for the following reference are represented as 000, 001, 010, and 011. The initial 0 in these positions signifies the strip. As described in our algorithm, the strip encapsulates the combination of compared sequences. In the example provided in this paper, where a single sequence is compared to another single sequence, only one strip (qubit) is needed. This strip represents either sequence 0 or sequence 1, thus serving as the data encoding component.

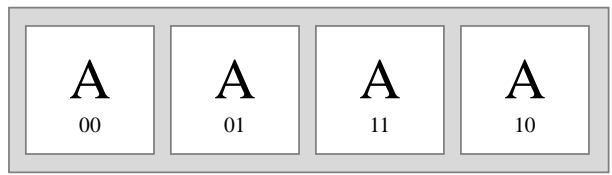

**Figure 6. A sequence built from four letters.** Each letter has its position, which is identified by the numbers below. We defined for each letter a quantum position. The letter "A" represents Adenine. To simplify the visualization of the results, we're using one sequence, *strip1* here, only one letter with four positions.

At *strip1*, the comparison sequence to the reference sequence *strip0*, the nucleic base T is taken, with the four possible positions, as illustrated in Figure 7.

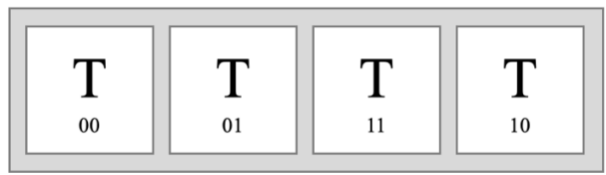

**Figure 7. The letter "T" represents Thymine**, which has the same numbers (positions) as Adenine.

The process flow of the similarity search between the two gene sequences is implemented in this experiment with a similar approach as in the publication of Yan et al. [41]. Figure 8 shows the scheme for parallel comparison of quantum images, which reflects the process flow of the experiment in this work.

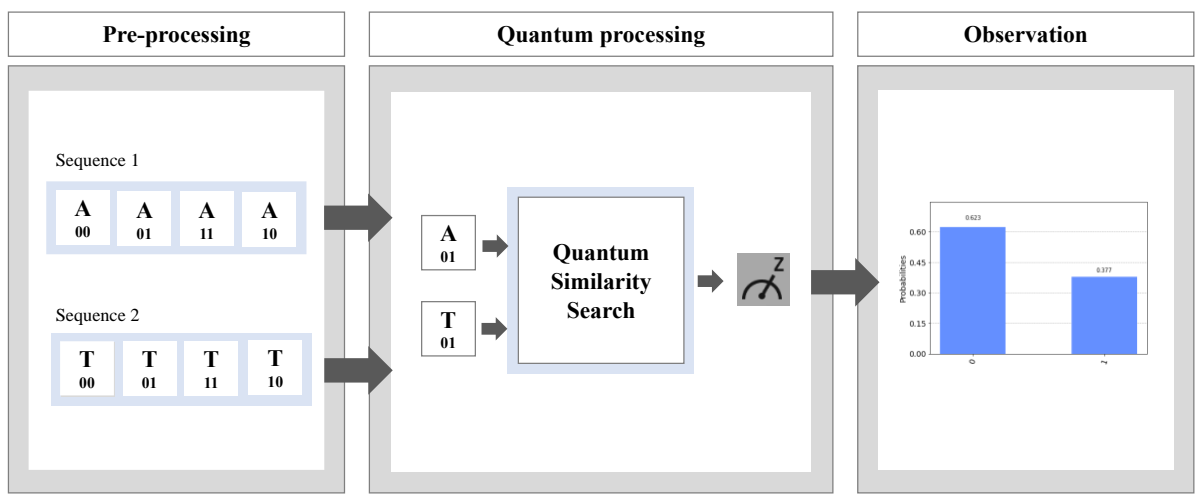

**Figure 8. A parallel comparison of two sequences**. The process is divided into three steps. They start with pre-processing to represent a sequence on a quantum computer. In the following quantum process, two sequences run a similarity search per letter. We compare two letters with one position (01) to simplify the visualization results. Ultimately, the measurement provides a histogram of whether the result between these two sequences is similar.

First, as shown in Figure 8, the pre-processing step generates a quantum circuit representing each gene sequence using amplitude estimation techniques. Then, both gene sequences are compared using the pairwise comparison method, which determines the rotation difference between each sequence. This process is completed by measuring the strip qubit, which generates a snapshot, or shot, of the resulting comparison. These shots are taken 8,000 times in the experiment. Finally, we view the result counts on a histogram. The result we are most concerned with is the probability of 1. This is the value we add as a parameter to determine the similarity score of the two sequences. To determine the similarity between the sequences, we must first extrapolate the probability results of 1, $P_1$. We then use the $P_1$ result value as a parameter to determine the similarity score between sequence1 and sequence2 as shown in the following similarity equation:

$$\text{sim (sequence1, sequence2)} = 1 - 2P_1$$

Table 2 shows two gene sequences, columns 1 and 2, respectively, where each entry contains each specific value and their represented phase rotation angles. The differences between the four are shown in sequential order. The third column indicates the probability results of the state $|1\rangle$, $P_1$:

| Sequence 1 (Reference) | Sequence 2 (Compared) | Result probabilty of $\|1\rangle$, $P_1$ |
|---|---|---|
| (A) $\pi$ | (A) $\pi$ | 0 |
| (A) $\pi$ | (C) $\pi/2$ | 0,146 |
| (A) $\pi$ | (T) $\pi/6$ | 0,371 |
| (A) $\pi$ | (G) 0 | 0,5 |

**Table 2: Results of the two compared sequences**; comparing the two sequences on a quantum computer illustrate the varying probability results of $P_{1x}$.

**Identifying the similarity of the phase angles.** This section provides evidence and the experimental results obtained with the proposed quantum algorithm identifying sequence similarity by phase angles. The experiment results for the expectation value $P_1$ come from the probability output by measuring the strip qubit that connects the sequences to be compared. The expectation value $P_1$ in the above table represents the differences between the gene sequences for each position. An important role is played by the distance between the phases, defined by the phase angles in each position. The closer the two-phase angles in each position are, the smaller the expected value or $P_1$. Here $P_1$ is the expected value, sequence$_1$ is the reference sequence, and sequence$_2$ is the comparison sequence. After the measurement, the first thing that is determined is the probability of $P_1$. Using the probability of $P_1$, the similarity score *sim* (sequence1, sequence2) between the reference and comparison sequences is determined. The result from the *sim* equation indicates whether a change is present. If the similarity value =1, then the two sequences are the same, whereas if the similarity value is less than 1, there is a difference between the two sequences.

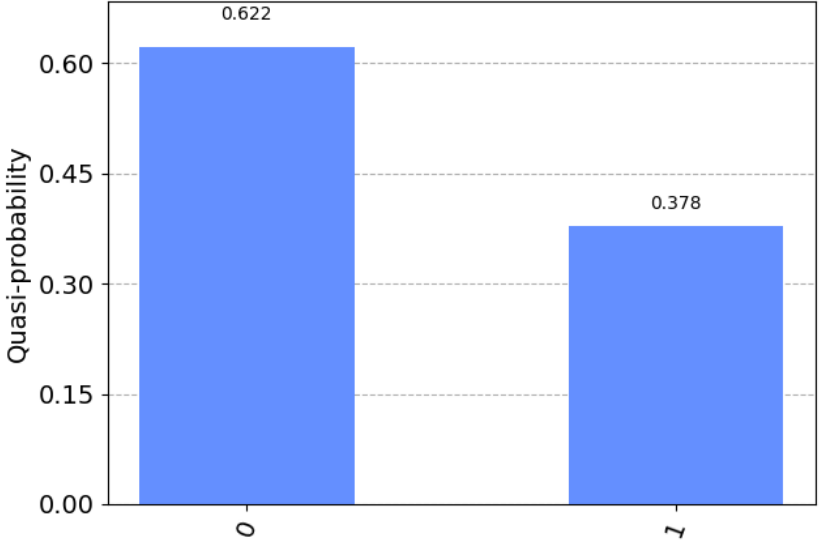

**Figure 9. Measured results of the strip qubit;** after running 8000 shots, the value for $|1ñ$ $P_1$ is 0.378.

In this case, the probability of getting a |1⟩ state result is 0.378, as shown in Figure 9. We use this result to include in the similarity equation to determine the similarity between the two sequences. This results in the following calculation for the similarity score:

$$\text{sim (sequence1, sequence2)} = 1 - (2 * 0.378) = 0.246$$

It results in the following interpretation: A similarity score of approx. 24.6% that both gene sequences are identical, indicating a differentiation between the two sequences.

## 3 Discussion

The goal of this work was two-fold. First, to represent a gene sequence as a quantum state based on FRQI. Second, to perform a comparison, utilizing the differences between phase angles of the two gene sequences and calculating the similarity score between them. It allowed us to illustrate that a quantum system makes sequential search possible.

However, due to noise and errors, current quantum systems' limitations made it impossible to perform a similarity score with an entire gene sequence that can contain millions upon millions of values. It is known that the human genome has about 3 billion base pairs, which is not currently possible to map completely as a quantum state. Nevertheless, this project has demonstrated that quantum computers have the potential to solve complex problems such as similarity scoring faster and with less memory in principle than classical computers. Because comparing two sequences of the size 100.000 with a 4-bit Integer will end in a memory consumption of 37 gigabytes [22,42]. The Hirschberg algorithm reduces the memory consumption to a linear space of O(n) [26,43]. With Quantum memory, information can be stored as a qubit superposition. The entanglement of qubits results in an exponential growth of memory. A one-dimensional sequence requires only ($[\log_2 n]$) qubits and N bits in a classical system [19,27,44]. The memory consumption is ($[\log_2 n + \log_2 m]$) qubits for comparing two sequences.

The approach in this work could be used on larger quantum computers in the future, expanded with more base pairs, and even analyzed with multiple gene sequences in parallel. Moreover, this approach is just one of many others which could be used for a mutation search. Some methods could be string comparison using hamming distance [45], string comparison using Grover's search algorithm (*35*), or, as described in the article by Niroula, P., and Nam, Y. This quantum pattern-matching algorithm matches a search string (pattern) of length M inside a longer text of length N [47].

Quantum computers are still in the early stages and are subject to several challenges. Quantum systems have potential use in various applications in life sciences [48] or healthcare, where breakthroughs are expected soon. Quantum mechanical calculations should make it possible to quantitatively predict molecules' properties [49].

Another essential use case is the application in the area of genomics. Big Data analytics can analyze ever-larger data generated by wearables, inside content, and eHealth apps. In addition, genetic testing is also increasingly in demand. Quantum computing and faster DNA sequencing would enable more comprehensive analyses of this data and lead to a speedier diagnosis.

Because of the complexity, we see particular potential in using quantum technologies in systems medicine [50]. Diseases are complex, as we have painfully discovered in the recent pandemic. Not only the genetic code but also the microbiome, the proteome, the metabolome, or the virome may play a crucial role. All systems interact and generate a higher degree of complexity that we can hardly manage with classical computers [51]. Thus, we propose "quantum systems medicine" for further medical research.

Our approach's novelty is applying a similarity method to compare quantum images in genome sequencing using the FRQI framework. It allows for comparisons at a fine granularity, down to single letters or amino acids within gene sequences. This breakthrough enables us to capture and compare subtle genetic variations, providing a more precise examination of gene information than traditional methods.

In this article, we present algorithmic advantages regarding time and space efficiency and accuracy.

**Time**: The comparison process is fast and costless since it involves a single quantum gate without any control condition to simultaneously transform the entire information encoding the two quantum images. It eliminates the need for iterative calculations and reduces the computational time significantly. Moreover, the quantum measurement, which determines the quantum system's state, is performed only at the end, contributing to time savings.

**Space**: The encoding of the gene sequences is given, and the number of sequences remains constant throughout the comparison process. It eliminates concerns about the scale of the encoding, making it more efficient in terms of space utilization. The encoding with different bases of the gene ensures that the number of sequences being compared does not impact the space requirements, unlike traditional methods where memory consumption grows with the size of the sequences.

**Accuracy**: The article builds upon existing alignment approaches used in genome sequencing, such as string-matching algorithms for exact matches and the Levenshtein distance for approximate matches. These established algorithms provide accurate results and are well-established in the field. Using quantum memory and entanglement allows for precise and reliable comparisons between sequences.

In contrast to current state-of-the-art methods, which often involve high costs and time requirements, the proposed approach offers advantages in terms of time, space, and accuracy. The Needleman-Wunsch algorithm, for example, has a time complexity of $O(mn)$ and a memory consumption of $O(nm)$ for comparing two sequences of size 100,000. In comparison, our approach significantly reduces both time and memory requirements. With its exponential memory growth through qubit entanglement, Quantum memory offers a more efficient representation of sequences, requiring fewer qubits than classical systems. The memory consumption for comparing two sequences is given by $(\lceil \log_2(n) \rceil + \log_2(m))$ qubits, ensuring efficient utilization of space resources.

In summary, our approach offers algorithmic advantages: reduced time complexity, improved space efficiency, and accurate sequence comparisons. The novelty lies in adapting the FRQI algorithm to perform similarity comparisons of quantum images in genome sequencing, enabling examination at the individual letter or amino acid level. This breakthrough enhances the accuracy and resolution of sequence comparisons, opening new possibilities in biological data analysis.

# 4 Materials, methods, and limitations

**Limitations of quantum computers.** The conducted investigation in this domain is subject to certain restrictions due to the limitations of current quantum computers. Data analysis with a real dataset could not be performed due to the circuit depth necessary to represent the data. Therefore, the similarity comparison in this work is limited to two short gene sequences. Each gene sequence has four nucleotides, primarily used as an example that can, over time, scale as the technology of quantum systems continues to evolve. There are three main challenges:

**1. Scalability**: at the time of this writing, there are 433-qubit machines with short decoherence times. We used the 5-qubit machine in this study because only a subset of a gene sequence containing four nucleotides was examined.

**2. Error rates:** if the circuit is too deep, that is, too many quantum operators are used and the coherence time of the qubits is exceeded, the results will contain some noise which affects the precision of the results. This is where a higher quantum volume may help in the future as we get to error correction and, eventually, logical qubits.

**3. Data complexity:** because a single gene already contains many nucleotides (sequences of letters), only a part of a gene section can be replicated due to hardware limitations. This is because there are several quantum operators behind a base pair, which represent the complexity. In that case, the whole gene could be mapped on the quantum computer, and a similarity comparison for multiple gene sequences could be used with more powerful quantum computers.

**Future work.** Gene sequencing and analysis is a critical yet complex step in medical research, where the genes are still not fully understood. Scientists are researching and assigning the respective functions of the individual genetic building blocks. An analysis of genes represents only a tiny part of the overall complexity of a disease. Many different components are involved in holistic systems medicine, such as studying proteins, small molecules, chemical reactions, bacteria, viruses, or even the social network in which people interact. Although complexity is increased many times [51], combining natural language processing (NLP) and quantum computing could lead to new insights. NLP, part of artificial intelligence, can be applied to image and text analysis. NLP offers the possibility of performing an otherwise very time-consuming investigation much faster and examining more significant amounts of textual data. An example application is the Artificial Intelligence Double Funnel by Buchkremer et al. [52].

And it is at this point that quantum computers could be of great help in analyzing these complex systems. As the increase in performance of quantum systems become available, more complex tasks could be solved with the help of quantum computers, which could include decoding the human genome. The processor architectures and performance, such as hardware quality and continuing development in error mitigation and error correction, are continuously improving.
IBM plans a 1,121-qubit quantum computer for 2023 and many error suppression and mitigation techniques, which will significantly advance and enable high scaling [53]. Even though we are still at the beginning of a very long road, quantum computers will open many possibilities for us in the future that will help solve currently intractable problems.

**Materials and Methods.** In this section, we want to provide more details about implementing the algorithm described in this work. The comparison process consists of the following steps:

**Step 1:** Assign each element a rotation angle; in this case, we used those defined in Table 1.

**Step 2:** Create a quantum circuit that includes several qubits that will represent the *index* number of the characters in each sequence string as a value of logN, where N is the number of characters in each sequence. In this example, we use N=4. Therefore, log4 = 2 qubits. Then add the strip qubit; in this case, we will use a single qubit to represent the strip since we are comparing two sequences. And finally, a single qubit encodes the value of the sequence element. Finally, add a classical register to read the results of the strip qubit. Table 3 illustrates the assignment labels.

| Basic Circuit | Definition |
| --- | --- |
| Strip 0 | References the gene sequence |
| Idx 0 | Position (X-axes) |
| Idx 1 | Position (Y-axes) |
| Dna 0 | Represents the four nuclear bases, each with phase angles |

**Table 3. Circuit labels and definitions**. Each qubit(s) is identified to illustrate whether the qubit applies to the position or the value.

**Step 3:** Set the strip and index qubits into a superposition state using Hadamard gates, as illustrated in Figure 4.

**Step 4:** Encode each strip, index, and sequence value using a multi-control, single-target gate, as illustrated in Figures 4 and 5. In this case, the gate will consist of three controls and one target, where the controls align with the strip and index qubits, and the target gate is a rotation gate aligned with the value of the sequence (either A, C, T, G), labeled DNA in the circuit.

**Step 5:** Set the control to capture each index position of each strip, denoted in Figures 4 and 5.

**Step 6:** Add another Hadamard gate to the strip qubit to complete the circuit.

**Step 7:** Measure the strip qubit; this will be used to determine the difference between sequences based on the result of $P_1$. The basic circuit is defined as follows and represented in Figures 4 and 5.

We now present an application of a quantum algorithm based on the Flexible Representation of Quantum Images (FRQI) applied to biological sequences. In its seminal definition, this algorithm provides a quantum representation of images that allows efficient encoding of classical data into a quantum state, i.e., color information and pixel position. Within this algorithm, encoding classical data into a quantum state requires a polynomial number of simple gates.

The idea here is to leverage the quantum encoding techniques to represent the varying nucleotide (4) (or amino acid - 20) representations using the Multi-Control-RY gate (MCRY).

Here we define the quantum state |Sequence (θ)ñ as a normalized state that encodes with this formulation the genetic sequences to compare, as a function of q:

$$|Sequence(\theta)\rangle = \frac{1}{2^n} \sum_{i=0}^{2^{2n}-1} (\cos\theta_i |0\rangle + \sin\theta_i |1\rangle) \otimes |i\rangle$$

A simple example of a 4-character nucleotide sequence, where the quantum registers represent the strip, index (reference in the gene sequence), and the nucleotide basis value, is given below, with corresponding θ angles (nitrogen bases) and associated Kets (position encoding):

$$|Sequence1\rangle = \frac{1}{2}[(\cos\theta_0|0\rangle + \sin\theta_0|1\rangle) \otimes |000\rangle + (\cos\theta_1|0\rangle + \sin\theta_1|1\rangle) \otimes |100\rangle$$
$$+ (\cos\theta_2|0\rangle + \sin\theta_2|1\rangle) \otimes |101\rangle + (\cos\theta_3|0\rangle + \sin\theta_3|1\rangle) \otimes |111\rangle$$
$$+ (\cos\theta_4|0\rangle + \sin\theta_4|1\rangle) \otimes |010\rangle + (\cos\theta_5|0\rangle + \sin\theta_5|1\rangle) \otimes |011\rangle$$
$$+ (\cos\theta_6|0\rangle + \sin\theta_6|1\rangle) \otimes |001\rangle + (\cos\theta_7|0\rangle + \sin\theta_7|1\rangle) \otimes |110\rangle]$$

In this example, we have set all units of the first sequence (theta 1) to p/4 (A) and the second sequence (theta2) to p/24 (T) and obtained a probability of P1 to 0.378, which resulted in a similarity score of 0.246, or 24.6%.

## Acknowledgments
M.G. is supported by CERN through CERN Quantum Technology Initiative.

## Data Availability Statement
We used the 5-qubit IBM cloud machine in this study because only a subset of a gene sequence containing four nucleotides was examined. This work was performed based on the Qiskit textbook chapter on quantum image processing (available at [https://learn.qiskit.org/course/ch-applications/flexible-representation-of-quantum-images-frqi](https://learn.qiskit.org/course/ch-applications/flexible-representation-of-quantum-images-frqi)).

## Funding
This project was not funded.

## Author contributions
The authors of this paper contributed the following:
R. B. conceived the idea of searching genetic sequences as letters on quantum computers. R. B. and B.K.K. conceived the study. R. L. identified the initial concept to implement this work based on the Qiskit textbook chapter on quantum image processing for a bioinformatic application. M. G. contributed to the coding of the sequences using the methods described above. J. B. contributed her research in the field of bioinformatics, providing the context of gene sequencing and biological sequence analysis. C. B. contributed his research in DNA sequencing with a particular focus on the performance and memory consumption of traditional alignment methods such as BLAST and FASTA. B.K.K., who authored her graduate thesis on this topic and served as the motivation to write this paper. R. B. served as B.K.K.´s graduate advisor on her graduate thesis at the FOM University of Applied Sciences. All authors reviewed and approved the final manuscript.

## Competing interests
The authors declare that they have no competing interests.

## Additional Information
All data needed to evaluate the conclusions in the paper are present in the article and/or the Supplementary Materials.